\documentclass[twocolumn,times]{aastex61}
\usepackage{natbib}

\newcommand{\kms}{km\,s$^{-1}$}
\newcommand{\cii}{[\ion{C}{2}]}
\newcommand{\ci}{[\ion{C}{1}]}

\newcommand{\lfir}{$L_{\mathrm{FIR}}$}
\newcommand{\ltir}{$L_{\mathrm{TIR}}$}
\newcommand{\lcii}{$L_\mathrm{[CII]}$}

\newcommand{\lci}{$L_\mathrm{[CI]}$}

\newcommand{\lsun}{$L_\sun$}
\newcommand{\msun}{$M_\sun$}

\newcommand{\msunyr}{$M_\sun$\,yr$^{-1}$}

\received{2017 August 29}
\revised{2017 October 13}
\accepted{2017 October 17}

\shorttitle{}
\shortauthors{Venemans et al.}

\begin{document}

\title{Copious Amounts of Dust and Gas in a $z=7.5$ Quasar Host Galaxy}

\correspondingauthor{Bram P.\ Venemans}
\email{venemans@mpia.de}

\author{Bram P.\ Venemans}
\affiliation{Max-Planck Institute for Astronomy, K{\"o}nigstuhl 17, D-69117
  Heidelberg, Germany}

\author{Fabian Walter}
\affiliation{Max-Planck Institute for Astronomy, K{\"o}nigstuhl 17, D-69117
  Heidelberg, Germany}
\affiliation{Astronomy Department, California Institute of Technology, MC105-24, Pasadena, CA 91125, USA}
\affiliation{National Radio Astronomy Observatory, Pete V. Domenici Array Science Center, P.O. Box 0, Socorro, NM 87801, USA}

\author{Roberto Decarli}
\affiliation{Max-Planck Institute for Astronomy, K{\"o}nigstuhl 17, D-69117
  Heidelberg, Germany}
\affiliation{Osservatorio Astronomico di Bologna, via Gobetti 93/3, I-40129 Bologna, Italy}

\author{Eduardo Ba{\~n}ados}
\affiliation{The Observatories of the Carnegie Institution for Science, 813 Santa Barbara Street, Pasadena, CA 91101, USA}

\author{Chris Carilli}
\affiliation{National Radio Astronomy Observatory, Pete V. Domenici Array Science Center, P.O. Box 0, Socorro, NM 87801, USA}
\affiliation{Astrophysics Group, Cavendish Laboratory, JJ Thomson Avenue, Cambridge CB3
0HE, UK}

\author{Jan Martin Winters}
\affiliation{Institut de Radioastronomie Millim\'etrique (IRAM), 300 rue de la Piscine, F-38406 Saint Martin d'H\`eres, France}

\author{Karl Schuster}
\affiliation{Institut de Radioastronomie Millim\'etrique (IRAM), 300 rue de la Piscine, F-38406 Saint Martin d'H\`eres, France}

\author{Elisabete da Cunha}
\affiliation{Research School of Astronomy and Astrophysics, Australian National University, Canberra, ACT 2611, Australia}

\author{Xiaohui Fan}
\affiliation{Steward Observatory, The University of Arizona, 933 North Cherry Avenue, Tucson, AZ 85721-0065, USA}

\author{Emanuele Paolo Farina}
\affiliation{Max-Planck Institute for Astronomy, K{\"o}nigstuhl 17, D-69117
  Heidelberg, Germany}

\author{Chiara Mazzucchelli}
\affiliation{Max-Planck Institute for Astronomy, K{\"o}nigstuhl 17, D-69117
  Heidelberg, Germany}

\author{Hans-Walter Rix}
\affiliation{Max-Planck Institute for Astronomy, K{\"o}nigstuhl 17, D-69117
  Heidelberg, Germany}

\author{Axel Wei\ss}
\affiliation{Max-Planck-Institut f{\"u}r Radioastronomie, Auf dem H{\"u}gel 69, D-53121 Bonn, Germany}

\begin{abstract}
We present IRAM/NOEMA and JVLA observations of the quasar J1342+0928 at $z=7.54$ and report detections of copious amounts of dust and \cii\ emission in the interstellar medium (ISM) of its host galaxy. At this redshift, the age of the universe is 690\,Myr, about 10\% younger than the redshift of the previous quasar record holder. Yet, the ISM of this new quasar host galaxy is significantly enriched by metals, as evidenced by the detection of the \cii\,158\,$\mu$m cooling line and the underlying far-infrared (FIR) dust continuum emission. To the first order, the FIR properties of this quasar host are similar to those found at a slightly lower redshift ($z\sim6$), making this source by far the FIR-brightest galaxy known at $z\gtrsim7.5$. The \cii\ emission is spatially unresolved, with an upper limit on the diameter of 7\,kpc. Together with the measured FWHM of the \cii\ line, this yields a dynamical mass of the host of $<$$1.5\times10^{11}$\,\msun. Using standard assumptions about the dust temperature and emissivity, the NOEMA measurements give a dust mass of $(0.6-4.3)\times10^8$\,\msun. The brightness of the \cii\ luminosity, together with the high dust mass, imply active ongoing star formation in the quasar host. Using \cii--SFR scaling relations, we derive star formation rates of 85--545\,\msunyr\ in the host, consistent with the values derived from the dust continuum. Indeed, an episode of such past high star formation is needed to explain the presence of $\sim$$10^8$\,\msun\ of dust implied by the observations.  
\end{abstract}

\keywords{cosmology: observations --- galaxies: high-redshift --- galaxies: ISM --- galaxies: active}

\section{Introduction} 
\label{sec:introduction}

The advent of large, wide-area optical and infrared surveys has resulted in the discovery of luminous quasars out to the highest redshifts, $z\gtrsim7$ \citep[e.g.,][]{fan06a,ven13,ban16,maz17b}. These quasars are powered by supermassive, $\sim$$10^9$\,\msun\ black holes that accrete near the Eddington limit \citep[e.g.,][]{der14,maz17b}. Since their discovery, the presence of such massive black holes has been a puzzle, as they require either very efficient accretion mechanisms from stellar black hole seeds \citep[$\sim$100\,\msun, e.g.,][]{vol12}, or the formation of massive seeds, e.g.\ via direct gas collapse \citep[$\sim$$10^{3-5}$\,\msun; e.g.,][]{aga12,reg17}.

Likewise, studies of the {\em host galaxies} of these distant quasars have revealed the presence of large amounts of dust and gas out to $z\sim7$ \citep[e.g.,][]{ber03b,wal03,mai05,ven12,ven16,wan13,wil15}. The associated molecular gas masses are $>$$10^{10}$\,\msun\ and provide the fuel for long-lasting ($>$10$^{7-8}$\,years) episodes of ultra-luminous infrared galaxies (ULIRG)-like star formation (with star-formation rates (SFRs)\,$\sim$100--1000\,\msunyr). 

The observed tracers (\cii, \ci, CO, far-infrared (FIR) continuum) require that the interstellar medium (ISM) of the host galaxies is chemically enriched. While metal-enriched material is also evident from broad emission lines in the quasar's rest-frame UV spectrum \citep[e.g.,][]{jia07,der14}, these emission lines originate from a concentrated region ($\ll$1\,pc) around the black hole, the so-called broad-line region (BLR). However, since the BLR total mass is only $10^{4-5}$\,\msun\ \citep[e.g.,][]{ferl04}, a modest amount of metals, $\sim$$10^{3-4}$\,\msun, can explain the observed high BLR metallicities \citep[$Z\sim10\,Z_\sun$, e.g.,][]{die03,jua09}. On the other hand, the enrichment of the quasar host is on significantly larger scales \citep[$\sim$kpc; e.g.,][]{wal09b,wan13,ven16}, which requires a galaxy-wide chemical enrichment due to extended star formation. The enrichment on galactic scales seen in dust and gas thus puts constraints on early metal production in the quasar host \citep[e.g.,][]{mic10,gal11b}. Star formation is only indirectly traced by the FIR emission, and to date the stellar component of the quasar host remains elusive \citep[e.g.,][]{dec12}.

To further constrain the formation of dust and enrichment of gas in the ISM in the earliest galaxies, studies need to be pushed back in time, i.e., to the highest possible redshifts. The highest-redshift quasar in which gas and dust have been detected to date is J1120+0641 \citep{mor11} at a redshift of $z=7.09$ \citep{ven17a}. Here, we report the detection of gas and dust emission in a newly discovered quasar at $z=7.5$, J1342+0928 \citep{ban17a}. The redshift of the quasar derived from the \ion{Mg}{2} line is $z_\mathrm{MgII}=7.527\pm0.004$ (age of the universe: 690\,Myr). From the width of the \ion{Mg}{2} line and the strength of the continuum, \citet{ban17a} estimate that the quasar is powered by accretion onto a $7.8^{+3.3}_{-1.9}\times10^8$\,\msun\ black hole. The quasar has 
an absolute magnitude of $M_{1450\,\mathrm{\AA}}\!=\!-26.8$
and shares many of the physical properties seen in quasars observed at $z\sim6-7$ \citep{ban17a}. 

Throughout this Letter, we adopt a concordance cosmology with $\Omega_M=0.3$, $\Omega_\Lambda=0.7$, and $H_0=70$\,\kms\,Mpc$^{-1}$. The physical scale at $z=7.54$ is 5.0\,kpc\,arcsec$^{-1}$. All magnitudes are on the AB system.

\section{Observations}

\subsection{NOEMA Observations}
\label{sec:noema}

\cii$_{3/2-1/2}$\,158\,$\mu$m (hereafter \cii), CO(7--6), CO(10--9), H$_2$O, and \ci$_{2-1}$ observations of J1342+0928 were
performed with the IRAM NOrthern Extended Millimeter Array (NOEMA). Observations were done with the array in compact configuration, using 7--8 antennas. All of the NOEMA data have been reduced using the latest version of the GILDAS software\footnote{http://www.iram.fr/IRAMFR/GILDAS}.

The observations were gathered between 2017 March 15 and May 21 in various visits. For the \cii\ observations, the NOEMA receiver 3 (1.2\,mm) was tuned to 224.121\,GHz in the first execution, and to 222.500\,GHz in all the other visits, in order
to better encompass the line within the WideX 3.6\,GHz bandwidth. The CO(10--9) line and the H$_2$O 3(2,1)-3(1,2) line at rest-frequency 1162.91\,GHz were observed in a single frequency setting, with NOEMA receiver 2 (2\,mm) tuned to 135.495\,GHz. The CO(7--6) and \ci$_{2-1}$ lines were observed with the 3\,mm receivers tuned to 94.587\,GHz. The radio quasar 1345+125 served as amplitude and phase calibrator. Additional calibrators used in the bandpass calibration included 3C273 and
3C454.3. The star MWC~349 was used to set the absolute flux scale. Measured line fluxes and continuum flux densities in Section~\ref{sec:results} and Table~\ref{tab:res} only include statistical errors and do not take the systematic flux calibration uncertainties of $\sim$10\% into account. The total integration time on-source was 13.6, 3.8, and 11.1\,hr (8 antenna equivalent) in the 1\,mm, 2\,mm, and 3\,mm bands, respectively. Imaging was performed using natural weighting, in order to maximize sensitivity. The resulting synthesized beams are 2\farcs4$\times$1\farcs5, 3\farcs6$\times$2\farcs5, and 5\farcs8$\times$3\farcs4 and the final 1\,mm, 2\,mm, and 3\,mm cubes reach a sensitivity of 0.47\,mJy\,beam$^{-1}$, 0.41\,mJy\,beam$^{-1}$, and 0.17\,mJy\,beam$^{-1}$ per 100\,\kms\ channel (1-$\sigma$), respectively. In the 1\,mm cube, both the \cii\ emission and the underlying dust continuum are significantly detected (Figure~\ref{fig:spectrum} and Section~\ref{sec:results}), while no emission was detected in the other two cubes. 

In the continuum images, an additional source is located $\sim$10\arcsec\ northeast of the quasar (see Figure~\ref{fig:maps}) with flux densities of $S_{223.5\,\mathrm{GHz}}=434\pm73$\,$\mu$Jy, $S_{135.5\,\mathrm{GHz}}=197\pm46$\,$\mu$Jy, and $S_{95\,\mathrm{GHz}}=41\pm16$\,$\mu$Jy. The spectrum of this object does not show emission lines. While the redshift remains unknown, the lack of line emission in the 1\,mm datacube, which covers a \cii\ redshift of $\Delta z \approx 0.1$ around that of the quasar, makes it unlikely that this source is physically associated with J1342+0928. 

\subsection{Very Large Array (VLA) Observations}

We searched for CO(3--2) emission from J1342+0928 with the VLA in 2017 April. The redshift of the source places the line at 40.4852\,GHz. The data also provided a deep continuum observation at 41\,GHz. A total of 9\,hr (7\,hr on-source), was spent using the 8\,bit, 2\,GHz bandwidth correlator mode for highest line sensitivity. An additional 3\,hr was spent using 3\,bit, 8\,GHz bandwidth from 40 to 48\,GHz for an additional continuum measurement.

Standard phase and amplitude calibration was performed, using J1331+305 to set the absolute gain scale and bandpass, and J1347+122 to determine complex gains as a function of time. Phase stability was excellent.

The line data were imaged using natural weighting and smoothed to a velocity resolution of 44.5\,\kms. The synthesized beam is 2\farcs2$\times$2\farcs0, and the rms noise per channel was 0.10\,mJy\,beam$^{-1}$. We also created a 41.0\,GHz continuum image by suitably combining all the data. The rms noise of this continuum image is 5.7\,$\mu$Jy\,beam$^{-1}$. No line was found, but a potential continuum source is reported (Section~\ref{sec:cont}).

\section{The Host Galaxy of J1342+0928 at $z=7.5$}
\label{sec:results}

\begin{table}[!t]
\begin{center}
\caption{Observed and Derived Properties of J1342+0928 \label{tab:res}}
\begin{tabular}{lc}
\hline
\hline
R.A.\ (J2000) & 13$^\mathrm{h}$42$^\mathrm{m}$08$.\!\!^\mathrm{s}$097 \\
Decl. (J2000) & $+$09$^\circ$28$^\prime$38\farcs28 \\
$z_\mathrm{[CII]}$ & 7.5413$\pm$0.0007 \\
$F_\mathrm{[CII]}$ (Jy\,\kms) & 1.25$\pm$0.17 \\
FWHM$_\mathrm{[CII]}$ (\kms) & 383$\pm$56~ \\
EW$_\mathrm{[CII]}$ ($\mu$m) & 1.73$\pm$0.43 \\
$S_{223.5\,\mathrm{GHz}}$ ($\mu$Jy) & 415$\pm$73~ \\
$S_{135.5\,\mathrm{GHz}}$ ($\mu$Jy) & $<$139 \\
$S_{95\,\mathrm{GHz}}$ ($\mu$Jy) & $<$48 \\
$S_{41\,\mathrm{GHz}}$ ($\mu$Jy) & 15.0$\pm$5.7~ \\
$S_{1.4\,\mathrm{GHz}}$ ($\mu$Jy) & $<$432 \\
$F_\mathrm{CO(3-2)}$ (Jy\,\kms) & $<$0.081 \\
$F_\mathrm{CO(7-6)}$ (Jy\,\kms) & $<$0.13 \\
$F_\mathrm{CO(10-9)}$ (Jy\,\kms) & $<$0.32 \\
$F_\mathrm{[CI]}$ (Jy\,\kms) & $<$0.14 \\
$F_{\mathrm{H}_2\mathrm{O}, 1172\,\mathrm{GHz}}$ (Jy\,\kms) & $<$0.30 \\
$F_{\mathrm{H}_2\mathrm{O}, 1918\,\mathrm{GHz}}$ (Jy\,\kms) & $<$0.33 \\
\hline
\lfir\ (\lsun) & $(0.5-1.4)\times10^{12}$ \\
\ltir\ (\lsun) & $(0.8-2.0)\times10^{12}$ \\
\lcii\ (\lsun) & $(1.6\pm0.2)\times10^{9~}$ \\
\lci\ (\lsun) & $<$$7.8\times10^{7~}$ \\
$L^\prime_\mathrm{CO(3-2)}$ (K\,\kms\,pc$^2$) & $<$$1.5\times10^{10}$ \\
SFR$_\mathrm{TIR}$ (\msunyr) & 120--300 \\
SFR$_\mathrm{[CII]}$ (\msunyr) & ~~85--545 \\
$M_d$ (\msun) & $(0.6-4.3)\times10^8$ \\
$M_{C^+}$ (\msun) & ~~$4.9\times10^{6~}$ \\
$M_{H_2}$ (\msun) & $<$$1.2\times10^{10}$ \\ 
\hline
\end{tabular}
\end{center}
\end{table}

Our NOEMA observations reveal the gas and dust present in the host galaxy of J1342+0928. In Figures~\ref{fig:spectrum} we show the spectrum of the \cii\ emission line and the underlying dust continuum. A summary of the measurements is given in Table~\ref{tab:res}. 

\subsection{Far-infrared Luminosity and Implied Dust Mass}
\label{sec:cont}

\begin{figure}
\includegraphics[width=\columnwidth]{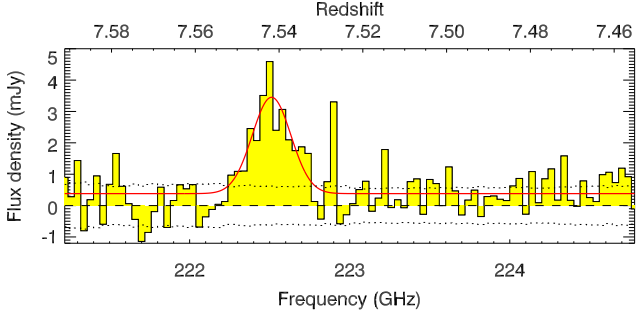}
\caption{NOEMA spectrum of the redshifted \cii\ emission line and the underlying continuum in J1342+092, extracted from the peak pixel in the datacube. The bin size is 40\,MHz, which corresponds to $\sim$54\,\kms. The dotted lines indicate +$\sigma$ and $-$$\sigma$, with $\sigma$ being the noise in each bin. The red, solid line is a flat continuum plus Gaussian fit to the spectrum (the fit values are reported in Table~\ref{tab:res}).}
\label{fig:spectrum}
\end{figure}

\begin{figure*}
\includegraphics[width=\textwidth]{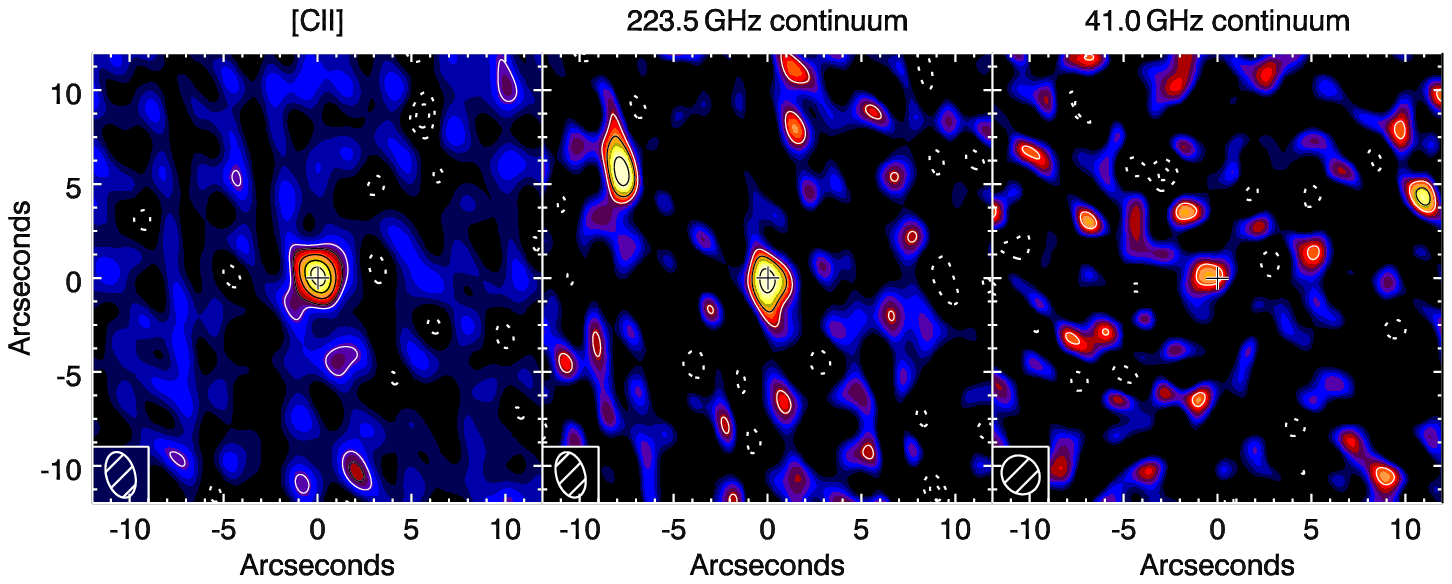}
\caption{Maps of the \cii\ emission (left), the continuum emission at 223.5\,GHz (middle), and the continuum at 41.0\,GHz of J1342+0928. The \cii\ map was created by averaging the continuum-subtracted datacube over 455\,\kms\ ($2.8\times\sigma_\mathrm{[CII]}$). The cross indicates the near-infrared position of the quasar. The beam is overplotted in the bottom left corner of each map. The contours show the emission at levels $-3$$\sigma$ and $-2$$\sigma$ (dotted lines) and +2$\sigma$, +3$\sigma$, +5$\sigma$, +7$\sigma$, and +9$\sigma$ (solid lines), where the $\sigma$ denotes the noise in the image (247\,$\mu$Jy\,beam$^{-1}$, 73\,$\mu$Jy\,beam$^{-1}$, and 5.7\,$\mu$Jy\,beam$^{-1}$, respectively). The nearby millimeter continuum source (Section~\ref{sec:noema}) can be seen toward the northeast in the middle panel.}
\label{fig:maps}
\end{figure*}

The dust continuum around the redshifted \cii\ emission (rest-frame wavelength of $\sim$158\,$\mu$m) has been detected at a signal-to-noise ratio (S/N)\,$\sim$\,6 and a strength of $S_{223.5\,\mathrm{GHz}}=415\pm73$\,$\mu$Jy (Figures~\ref{fig:spectrum} and \ref{fig:maps}). The source is not resolved with the 2\farcs4$\times$1\farcs5 (12.1$\times$7.3\,kpc$^2$) beam. We also estimated the source size in the {\it uv} plane and derive a source radius $<$0\farcs5, which is consistent with the size measurement of the continuum image. The position of the quasar host, R.A.\,=\,13$^\mathrm{h}$42$^\mathrm{m}$08$.\!\!^\mathrm{s}$097; decl.\,=\,+09$^\circ$28\arcmin38\farcs28, is consistent with the near-infrared location of the quasar \citep{ban17a}. The host galaxy is not detected in continuum in the other NOEMA setups down to 3\,$\sigma$ continuum limits of $S_{135.5\,\mathrm{GHz}}<139$\,$\mu$Jy and $S_{95\,\mathrm{GHz}}<48$\,$\mu$Jy. The VLA continuum map shows a potential source ($S_\mathrm{41\,GHz}=15.0\pm5.7$\,$\mu$Jy\,beam$^{-1}$ and S/N\,$\sim$\,2.6; Figure~\ref{fig:maps}), located $\sim$0\farcs7 from the \cii\ emission of J1342+0928.

To compute the far-infrared (rest-frame 42.5--122.5\,$\mu$m) and total infrared (TIR; 8--1000\,$\mu$m) luminosities, \lfir\ and \ltir, and the dust mass $M_d$ in J1342+0928, we follow the same procedure as outlined in \citet[][]{ven16}. In summary, we utilize three different models to estimate dust emission: a modified black body (MBB) with a dust temperature $T_d=47$\,K and an emissivity index of $\beta=1.6$ \citep[e.g.,][]{bee06} and two templates of local star-forming galaxies (Arp220 and M82) from \citet{sil98}. We also take the effect of the cosmic microwave background (CMB) on the dust emission into account \citep[e.g.,][]{dac13,ven16}. The mass of dust is derived both by assuming an opacity index of $\kappa_\lambda=0.77 (850 \mu\mathrm{m}/\lambda)^\beta$\,cm$^2$\,g$^{-1}$ \citep{dun00} and from scaling the Arp220 and M82 templates \citep{sil98}. We stress that due to the unknown shape of the dust continuum, the FIR and TIR luminosities remain highly uncertain, while the SFR and dust mass we derive crucially depend on the applicability of local correlations to this high-redshift source.

Scaling the NOEMA continuum detection of $S_{223.5\,\mathrm{GHz}}=415\pm73$\,$\mu$Jy to the three dust spectral energy distribution (SED) models results in luminosities of \lfir\,=\,$(0.5-1.4)\times10^{12}$\,\lsun\ and \ltir\,=\,$(0.8-2.0)\times10^{12}$\,\lsun. The derived dust mass is $M_d=(0.6-4.3)\times10^8$\,\msun. Applying the local scaling relation between \ltir\ and SFR from \citet{mur11} and assuming the infrared luminosity is dominated by star-formation \citep[e.g.,][]{lei14} results in an SFR of $120-300$\,\msunyr. This is significantly lower than the SFR derived for some of the quasar hosts at $z\sim6$ \citep[e.g.,][]{wal09b}, but very similar to the SFR in J1120+0641 at $z=7.1$ \citep{ven17a}. 

\subsection{Tentative Radio Continuum Emission}

\begin{figure}
\includegraphics[width=\columnwidth]{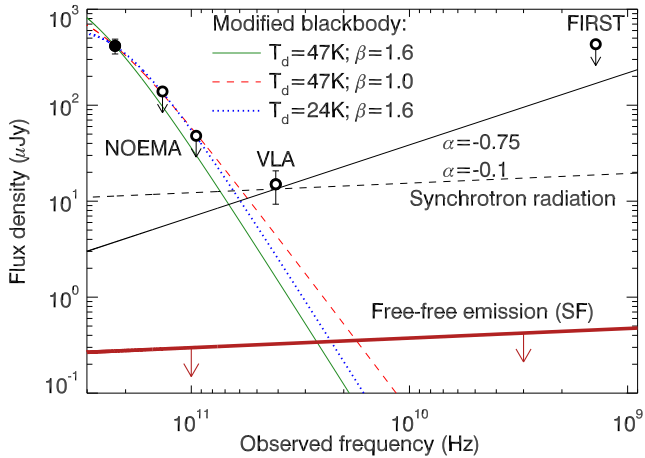}
\caption{Far-infrared and radio spectral energy distribution of J1342+0928. The data points from left to right represent the NOEMA 1, 2, and 3\,mm observations, the tentative VLA 41.0\,GHz detection, and the FIRST upper limit. Overplotted are three different dust SEDs scaled to the 1\,mm detection and two power laws describing radio synchrotron radiation. The dust model with canonical values ($T_d=47$\,K and $\beta=1.6$) agrees well with the upper limits on the continuum emission at 2 and 3\,mm, but predicts a much lower continuum flux density at 41.0\,GHz. A shallower $\beta$ ($\beta=1.0$) or a lower dust temperature ($T_d=24$\,K, slightly above the CMB temperature at $z=7.54$), illustrated by the dashed and dotted lines, also predicts a 41.0\,GHz flux density below that of the tentative VLA source. The upper limit in FIRST does not provide strong constraints on the slope of the radio emission.}
\label{fig:sed}
\end{figure}

We now look into the origin of the potential VLA continuum detection. The first possibility is that the source is spurious. The S/N is only 2.6, and as shown in Figure~\ref{fig:maps}, several positive noise peaks are visible close to the location of the quasar. It is therefore plausible that the 41\,GHz detection will disappear when adding more data. On the other hand, if the source is real, then the question is whether it is due to dust emission, free--free emission, or non-thermal processes, e.g., synchrotron radiation. The typical quasar dust SED, the MBB with $T_d=47$\,K and $\beta=1.6$ predicts flux densities of 90, 28, and 1.5\,$\mu$Jy at 135.5, 95, and 41.0\,GHz, respectively (Figure~\ref{fig:sed}). The limits in the NOEMA 2\,mm and 3\,mm bands are consistent with these expected flux densities, but the flux density measured in the VLA image is significantly ($\sim$10$\times$) higher than expected from the dust emission. A much shallower emissivity index ($\beta\ll1.5$) and/or a lower dust temperature, which would result in a higher flux density at 41\,GHz, can be ruled out by the nondetections at 135.5 and 95\,GHz (Figure~\ref{fig:sed}). Based on the derived SFR in the host galaxy (SFR\,$=85-545$\,\msunyr, Table~\ref{tab:res}), the strength of free--free emission at 41.0\,GHz is negligible \citep[$S_\mathrm{ff}\ll1$\,$\mu$Jy; e.g.,][]{yun02}. Alternatively, the flux density could be due to synchrotron radiation. We can estimate the radio loudness of the quasar using the 
radio-to-optical flux density ratio $R=S_{\mathrm{5\,GHz, rest}}/S_{\mathrm{4400\,\AA, rest}}$ with $S_{\mathrm{5\,GHz, rest}}$ and $S_{\mathrm{4400\,\AA, rest}}$ the flux densities at rest-frame 5\,GHz and 4400\,\AA, respectively \citep{kel89}. Assuming a radio continuum can be described by a power law ($f_\nu\propto\nu^\alpha$) with $\alpha=-0.75$ \citep[e.g.,][]{ban15a}, we derive $S_\mathrm{5\,GHz, rest}=363$\,$\mu$Jy. Following \citet{ban15a}, we derive $S_\mathrm{4400\,\AA, rest}=29$\,$\mu$Jy from the {\em WISE} W1 magnitude (W1\,=\,20.17). We obtain $R=12.4$, making
J1342+0928 a radio-loud quasar (where radio-loud is defined as $R>10$). Note that this is still consistent with the nondetection in the FIRST survey, with a 3\,$\sigma$ upper limit of $S_\mathrm{1.4\,GHz}<432$\,$\mu$Jy, as the expected flux density for J1342+0928 is $S_\mathrm{1.4\,GHz}\approx190$\,$\mu$Jy (Figure~\ref{fig:sed}). Deeper imaging at radio frequencies will provide a definitive answer.

\subsection{\cii\ Luminosity}
\label{sec:cii}

We detect the \cii\ emission line in J1342+0928 in the continuum-subtracted \cii\ map (Figure~\ref{fig:maps}) with an S/N\,$\sim$\,10. The spectrum, extracted from the peak pixel in the datacube, is shown in Figure~\ref{fig:spectrum}. From a Gaussian fit to the line, we derive a redshift of $z_\mathrm{[CII]}=7.5413\pm0.0007$, a line flux of $F_\mathrm{[CII]}=1.25\pm0.17$\,Jy\,\kms, and a dispersion of $\sigma_\mathrm{[CII]}=163\pm24$\,\kms\ (FWHM$_\mathrm{[CII]}=383\pm56$\,\kms); see Table~\ref{tab:res}. This corresponds to a \cii\ luminosity in this quasar of \lcii\,$=(1.6\pm0.2)\times10^9$\,\lsun, which is roughly $\sim$15\% brighter than J1120+0641 at $z=7.1$ \citep{ven17a} and a factor 3--5 fainter than the most \cii\ luminous quasar at $z\sim6$ \citep[e.g.,][]{mai05,wan13}. 

The redshift derived from the \cii\ line is higher than that derived from the UV emission lines of the quasar. The \ion{C}{4} and \ion{Mg}{2} lines are blueshifted by 6580$\pm$270\,\kms\ and 500$\pm$140\,\kms\ with respect to the \cii\ line. The \ion{Mg}{2} shift is close to the mean blueshift of the \ion{Mg}{2} line of 480\,\kms\ found in a sample of $z\sim6-7$ quasars \citep[e.g.,][]{ven16}. This could indicate the presence of an outflow \citep[e.g.,][]{maz17b}.

\begin{figure}
\includegraphics[width=\columnwidth]{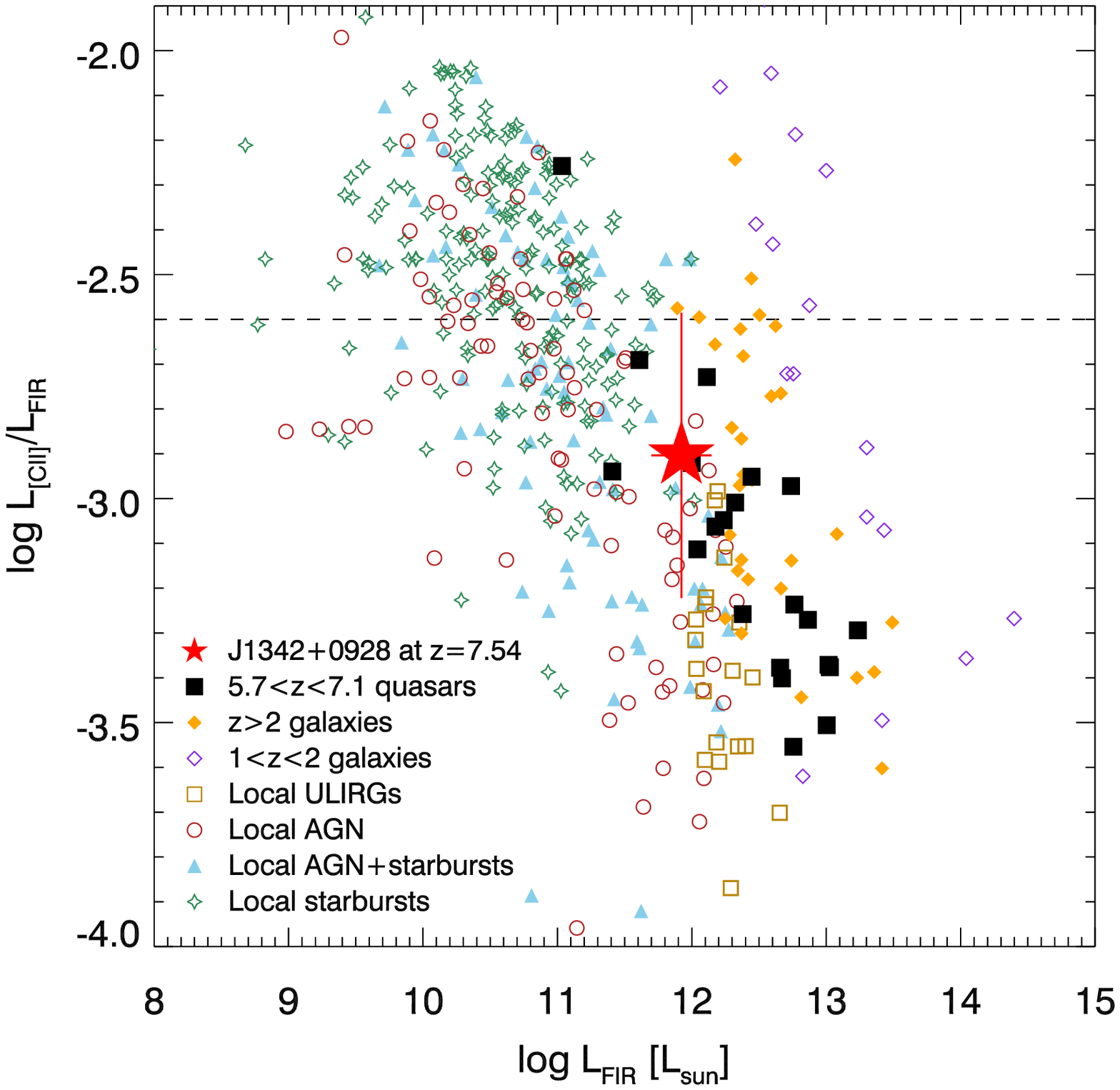}
\caption{\cii-to-FIR luminosity ratio vs.\ FIR luminosity. Plotted are values for starburst galaxies and active galactic nuclei (AGNs) in the local universe and at high redshift and for local ULIRGs \citep[][and references therein]{mai05,wal09b,dia13,ven16,maz17b}. The value for J1342+0928 is plotted as a red star. The dashed line indicates the median \lcii/\lfir\ ratio of local star-forming galaxies \citep{dia13}.}
\label{fig:lumrat}
\end{figure}

We measure a rest-frame \cii\ equivalent width of EW$_\mathrm{[CII]}=1.73\pm0.43$\,$\mu$m, which is consistent with the mean EW$_\mathrm{[CII]}$ of local starburst galaxies \citep[which have  $\langle$EW$_\mathrm{[CII]}\rangle=1.27\pm0.53$\,$\mu$m; see e.g.,][]{dia13,sar14} and higher than those of luminous ($M_{1450}<-27$) quasars at $z\sim6$ \citep[e.g.,][]{wan13}. The \cii-to-FIR luminosity ratio is \lcii/\lfir\,$=(0.6-2.6)\times10^{-3}$ (Figure~\ref{fig:lumrat}), again consistent within the large uncertainties with the \lcii/\lfir\ ratio of local star-forming galaxies that have a median \lcii/\lfir\,=\,$2.5\times10^{-3}$ \citep[e.g.,][]{dia13}. 

We can estimate the SFR from the \cii\ emission using the SFR--\lcii\ relations for high-redshift ($z>0.5$) galaxies of \citet{del14}: 

\begin{equation}
\mathrm{SFR}_\mathrm{[CII]}/\mathrm{M}_\sun\,\mathrm{yr}^{-1} = 3.0\times10^{-9} (L_\mathrm{[CII]}/\mathrm{L}_\sun)^{1.18}, 
\end{equation}

\noindent
with a systematic uncertainty of a factor of $\sim$2.5. With \lcii\,$=(1.6\pm0.2)\times10^9$\,\lsun\ we derive SFR$_\mathrm{[CII]}=85-545$\,\msunyr, which is similar to the SFR based on the TIR luminosity (Section~\ref{sec:cont}). 

The \cii\ emission is not resolved in the 2\farcs5$\times$1\farcs5 beam (Figure~\ref{fig:maps}). We fitted a 2D Gaussian to the \cii\ map using the CASA task ``imfit" and we derive a 1\,$\sigma$ upper limit on the size of 1\farcs7$\times$1\farcs2 (FWHM). A similar limit on the source diameter of $D<1\farcs0$ is found when fitting a 1D Gaussian to the {\it uv} data. This translates to an upper limit on the size of the \cii-emitting region of 8.4$\times$5.9\,kpc$^2$ or a diameter of $D\lesssim7$\,kpc. Approved observations with the Atacama Large Millimeter/ submillimeter Array (ALMA) at higher spatial resolution will put tighter constraints on the size of the host galaxy. 

From the strength of the \cii\ emission line, we can derive the mass of singly ionized carbon. In analogy to the formula to compute the mass of neutral carbon provided in \citet{wei05} and assuming optically thin \cii\ emission, the mass of singly ionized carbon can be calculated using

\begin{eqnarray}
M_{\mathrm{C}^+}/M_\odot &=& C m_\mathrm{C} \frac{8\pi k \nu_0^2}{h c^3 A} Q(T_\mathrm{ex}) \frac{1}{4} e^{91.2/T_\mathrm{ex}} L^\prime_\mathrm{[CII]} = \nonumber \\
&=&  2.92\times10^{-4} Q(T_\mathrm{ex}) \frac{1}{4} \mathrm{e}^{91.2/T_\mathrm{ex}} 
L^\prime_\mathrm{[CII]},
\label{eq:mcplus1}
\end{eqnarray}

\noindent
with $C$ the conversion between pc$^2$ and cm$^2$, $m_\mathrm{C}$ the mass of a carbon atom, $A=2.29\times10^{-6}$\,s$^{-1}$ the Einstein coefficient \citep{nus81}, $Q(T_\mathrm{ex})=2+4\mathrm{e}^{-91.2/T_\mathrm{ex}}$ the \ion{C}{2} partition function, and $T_\mathrm{ex}$ the excitation temperature. As \cii\ is emitted from the outer layers of photon-dominated region (PDR) clouds, $T_\mathrm{ex}\gtrsim100$\,K is a good assumption \citep[see, e.g.,][]{mei07}. Setting $T_\mathrm{ex}=100$\,K we derive $M_{\mathrm{C}^+}=4.9\times10^6$\,\msun. For $T_\mathrm{ex}=200$\,K (75\,K), the mass would be $\sim$20\% lower (higher). 

\subsection{Limits on the CO and \ci\ Luminosity}

We do not detect any of the other targeted emission lines in J1342+0928. To derive upper limits on the line fluxes, we averaged the datacubes over 2.8$\times\sigma_\mathrm{[CII]}$ (460\,\kms). We measured the following 3$\sigma$ upper limits: $F_\mathrm{CO(10-9)}<0.32$\,Jy\,\kms, $F_\mathrm{CO(7-6)}<0.13$\,Jy\,\kms, $F_\mathrm{[CI]}<0.14$\,Jy\,\kms, $F_\mathrm{CO(3-2)}<0.081$\,Jy\,\kms, and $F_{\mathrm{H}_2\mathrm{O,1172\,GHz}}<0.30$\,Jy\,\kms. 

The limits on the CO luminosity are $L^\prime_\mathrm{CO(10-9)}<5.2\times10^9$\,K\,\kms\,pc$^2$, $L^\prime_\mathrm{CO(7-6)}<4.3\times10^9$\,K\,\kms\,pc$^2$, and $L^\prime_\mathrm{CO(3-2)}<1.5\times10^{10}$\,K\,\kms\,pc$^2$. We can estimate a limit on the molecular gas mass $M_{H_2}$ by utilizing $M_{H_2}=\alpha L^\prime_\mathrm{CO(1-0)}$ with $\alpha$ the CO luminosity-to-gas mass conversion factor. Assuming the CO(3--2) emission is thermalized \citep[e.g.,][]{rie09}, the CO(1--0) luminosity is given by $L^\prime_\mathrm{CO(1-0)}=L^\prime_\mathrm{CO(3-2)}$. Adopting $\alpha=0.8$ \citep[e.g.,][]{dow98}, we set an upper limit on the molecular gas mass of $M_{H_2}<1.2\times10^{10}$\,\msun. 

The limiting luminosity of the \ci\ line is \lci\,$<7.8\times10^7$\,\lsun. With a measured \cii\ luminosity of \lcii\,=\,$(1.6\pm0.2)\times10^9$\,\lsun\ (Section~\ref{sec:cii}), we can set a lower limit to the \cii-to-\ci\ luminosity ratio of \lcii/\lci\,$>18$. Following \citet{ven17a}, we can compare this luminosity ratio to those predicted by the ISM models of \citet{mei07}. From the measured luminosity ratio we can exclude that the lines originate from a region where the X-ray radiation from the accreting black hole is dominating the emission. 

\subsection{Dynamical mass estimate}

From the velocity dispersion $\sigma$ of the \cii\ emission and the radius $R$ of the line emitting region, we can estimate a dynamical mass of the quasar host galaxy by utilizing the virial theorem: $M_\mathrm{dyn} = 3R\sigma^2/2G$ with $G$ as the gravitational constant. Assuming that the velocity dispersion can be derived from the Gaussian fit to the \cii\ emission (Figure~\ref{fig:spectrum}), and adopting a maximum radius of the \cii\ emission of $R<3.5$\,kpc (Section~\ref{sec:cii}), we infer a dynamical mass $M_\mathrm{dyn}<3.2\times10^{10}$\,\msun. If instead we assume that the \cii\ emission is in a rotating disk with inclination angle $i$ \citep[e.g.,][]{wan13,wil15,ven16}, we derive a higher dynamical mass of $M_\mathrm{dyn}<1.0\times10^{11}/\mathrm{sin}^2(i)$\,\msun. Adopting $i=55^\circ$, the median inclination angle of $z\sim6$ quasar hosts \citep{wan13}, the dynamical mass of J1342+0928 becomes $M_\mathrm{dyn}<1.5\times10^{11}$\,\msun, which is $\lesssim$190$\times$ higher that of the black hole \citep{ban17a}. To more accurately constrain the dynamical mass, high spatially resolved observations of the \cii\ emission are necessary.

\section{Concluding remarks}

We presented the detection of copious amounts of dust ($\sim$$10^8$\,\msun) and metal-enriched gas ($\sim$$5\times10^6$\,\msun\ of carbon in the singly ionized phase only) in a quasar host galaxy 690\,Myr after the Big Bang. The enrichment of the ISM in this source appears similar to other quasars at $z=6-7$  \citep[e.g.,][]{rie09,wan13,ven16} but much higher than what is typically found in non-quasar host galaxies at these redshifts \citep[e.g.,][]{wat15,pen16}.

We can only speculate which mechanism is responsible for the high mass in metals so early after the Big Bang. Because of the young cosmic age, asymptotic giant branch stars are thought to play only a marginal role \citep[e.g.,][]{mor03,jua09,gal11a}. On the other hand, type II supernovae (SNe) can produce significant amounts of dust, up to $\sim$1\,\msun\ per SN \citep[e.g.,][]{mat15}. For an initial mass function (IMF) similar to that of the Milky Way, the number of SNe is 1 per 200\,\msun\ of stars formed \citep[e.g.,][]{die06}. The implied stellar mass of J1342+0928 would then be $M_*=2\times10^{10}$\,\msun. Assuming a top-heavy IMF the implied stellar mass would be reduced by a factor of $\sim$3. In either case, such a massive stellar population should be easily detectable with the combined sensitivity, resolution, and wavelength coverage of the {\em James Webb Space Telescope} ({\em JWST}).

At these extreme redshifts, population III stars also provide a plausible enrichment mechanism. Metal-free stars with a mass $140<M$/\msun$<260$ could have dust yield as high as 15--30\% \citep[e.g.,][]{noz03}. Neglecting dust destruction, one would `only' require 2 million population III stars of 200\,\msun\ to create a mass of $10^8$\,\msun\ in dust, although the fast metal pollution may prevent the formation of so many population III stars \citep[e.g.,][]{mai10}.

The presented observations showcase how the study of quasar host galaxies at the highest redshifts can shed new light on the dawn of galaxy formation. Future ALMA and {\em JWST} observations will allow us to constrain the molecular gas mass, determine the shape of the FIR dust emission, and measure the size of the gas reservoir and to reveal the stellar population in the quasar host of this system.

\acknowledgments 
We thank the referee for providing valuable comments and suggestions. B.P.V., F.W., and E.P.F.\ acknowledge funding through the ERC grant ``Cosmic Dawn.'' 
Support for R.D.\ was provided by the DFG priority program 1573 ``The physics of the interstellar medium.'' We thank Amanda Karakas for help with estimating the metal production in stars. We are grateful to the JVLA and NOEMA for providing DDT observations. This work is based on observations carried our under project number E16AH with the IRAM NOEMA Interferometer. IRAM is supported by INSU/CNRS (France), MPG (Germany), and IGN (Spain).

\vspace{5mm}
\facilities{IRAM:Interferometer, EVLA}

\end{document}